# Charge carrier relaxation model in hopping system

Nianduan Lu[a], Ling Li[a*], Pengxiao Sun[a,b], and Ming Liu[a*]

Email: lingli@ime.ac.cn. liuming@ime.ac.cn.

[a] Laboratory of Nano-fabrication and Novel Devices Integrated Technology, Institute of Microelectronics, Chinese Academy of Sciences, Beijing 100029, People's Republic of China.

[b] Institute of Microelectronics, School of Physical Science and Technology, Lanzhou University, Lanzhou 730000, People's Republic of China.

**Abstract:**

The relaxation phenomena of charge carriers in hopping system have been demonstrated and investigated theoretically. An analytical model describing the charge carrier relaxation is proposed based on the hopping transport theory. The relation between the material disorder, electric field and temperature and the relaxation phenomena has been discussed in detail, respectively. The calculated results reveal that the increase of electric field and temperature can promote the relaxation effect in disordered systems, while the increase of material disorder will weaken the relaxation. The proposed model can explain well the stretched-exponential law by adopting the appropriate parameters. The calculation shows a good agreement with the experimental data for organic semiconductors.

# 1. Introduction

The relaxation phenomena in disordered systems, such as organic semiconductor and amorphous semiconductor, are of special importance for the research on charge carrier transport because the relaxation effect is caused essentially by the energy transition, diffusion process, free carrier mobility and lifetime, etc., and are in general characterized by some common patterns which are the presence of a slow dynamics, in spite of the large differences among them [1-5]. The stretched-exponential law was since proposed by Kohlrausch [6] and was widely used to describe the dielectric relaxation in polymers, a large number of models based on the stretched-exponential law, i.e., Förster direct-transfer model [7], hierarchically constrained dynamics model [8] and defect-diffusion model [9], were established and similar experiments had also been carried out to try to explain the relaxation phenomena in disordered systems [10, 11]. Unfortunately, the traditional phenomenological description of relaxation has no clear physical basis due to use the stochastic processes with time-independent transition rates leading to the usual exponential decay, and hence couldn't well interpret the intrinsic physical characteristic of the relaxation.

Several researchers attempted to introduce the energy, temperature and charge carrier transport to describe the relaxation phenomena, such as the multiple-trapping model [4, 12] and the Meyer-Neldel rule [13]. These models have a characteristic in common to introduce the transport energy concept (mobility edge) in the band tails, which had already been developed to describe hopping transport in an exponential band tail of disordered semiconductors. However, previous study revealed that the transport energy is only reasonable in the very low electric field and low carrier concentration regime [14]. Moreover, as has pointed out [15,16], electric field plays an important role for the charge carrier transport in organic semiconductors, and hence influences the relaxation effect, however, most of the relaxation models didn't take into account the electric field effect up to date.

It is generally agreed that charge carrier transport in disordered systems is via hopping between the molecule localized states. A large number of hopping transport theoretical investigations concerning charge carrier transport in disordered systems have been published [17, 18], and the progress in the analytical description of the relaxation phenomena has been made [11, 19, 20]. However, these models either derive the multiple-trapping model or try to artificially set the transport site and the transport route in a stochastic geometrical space in order to explain the property of hopping transport, which contrary to the nature of the hopping transport mechanism since the charge carriers in disordered systems are believed to be highly localized and the hopping of charge carriers via localized states is appropriated with the Gaussian energy distribution. The Gaussian density of states (DOS) will be assumed by the form, $g(E) = \frac{N_t}{\sqrt{2\pi}\sigma^*} \exp(-\frac{E^2}{2\sigma^{*2}})$, where $N_t$ is the number of states per unit volume, E is the energy, and $\sigma^* = \sigma/k_B T$ indicates the width of the DOS.

In the present work, we propose a procedure using hopping transport theory of charge carriers to describe the relaxation phenomena in disordered systems. The proposed model discards the traditional phenomenological description of relaxation and entirely based on the charge carrier transport in hopping systems expounds the relaxation. More important, our model for the first time introduces the electric field to explain the relaxation phenomena in disordered systems.

## 2. Model

Let us first consider the process of relaxation in disordered systems. It is assumed that the charge carriers rapidly moved down towards the deep energy by the hopping transport as soon as they entered into systems. Here, we ignored the downward movement time of charge carriers before the relaxation because the downward movement time of charge carriers is extremely short as compared with the relaxation time, and the starting time of relaxation phenomena was treated as t=0. The rate of the charge carrier hopping downward is given to a good approximation by the following expression:

$$v = v_0 \exp(-2\alpha R), \qquad (1)$$

where $\alpha$ is the inverse localization radius, $v_0$ is the attempt-to-escape frequency,

$R$ is the distance between two sites.

Due to the effect of hopping transport in disordered systems, the charge carriers will walk by hops from an initial site to the empty site at the closest range which will result in the occurrence of the current and energy decay. The time dependence of the current is determined by the time dependence of the concentration of charge carriers. Considering that the localized states are randomly distributed in energy and that space coordinates and the positions of these states, the localized states can be divided into fast state and slow state based on their different energy. The charge carriers will leave as fast as they arrive in the shallower states, while accumulated in the deeper states due to the slow leaving rates and also contained the charge carriers which happened to hop into the deeper states. The demarcation energy $E_d(t)$, separating the fast and slow states for the time dependent of energy, is defined as the energy at which the typical hopping-away rate $v_1$ is equal to the reciprocal of time [19]. At the same time, the demarcation energy $E_d(t)$ is also defined as the quasi-Fermi level $E_F$ of the disordered system at that moment [12, 21].

Since conduction is the result of a great deal of series of hops through the hopping space, one can derive the number of empty sites enclosed by the constant range R following the method in [22], as

$$N(T,\beta,E) = \frac{1}{8\alpha^3}\int_0^\pi d\theta \sin\theta \int_0^R dr 2\pi r^2 \int_{-\infty}^{R+E_s-r(1+\beta\cos\theta)} dE g(E)(1-f(E,E_F)), \quad (2)$$

where $\beta = Fe/(2\alpha k_B T)$, $F$ is the electric field, $k_B$ is the Boltzmann constant, T is the temperature, $f(E,E_F) = 1/(1+\exp(E-E_F))$ is the Fermi-Dirac distribution and $1-f(E,E_F)$ is the probability that the finial site is empty, θ is the angle between R and the electric field, $E_s$ is the energy of the initial localized state.

According to the variable range hopping theory, the mean hopping range could be obtained by setting $N(T,\beta,E) = 1$ in Eq. (2). By changing the integration variable, Eq. (2) could be divided into two parts which are corresponding to hopping upward

($\uparrow$) and hopping downward ($\downarrow$) of the charge carriers, respectively, can be written as

$$\begin{cases} (a) \ \dfrac{1}{24\alpha^3}\int_0^\pi d\theta \sin\theta \int_{E_s-R\beta\cos\theta}^{E_s+R} dEg(E)(1-f(E,E_F))(\dfrac{E_s-E+R}{1+\beta\cos\theta})^3 =1, & \uparrow \\ (b) \ \dfrac{R^3}{24\alpha^3}\int_0^\pi d\theta \sin\theta \int_{-\infty}^{E_s-R\beta\cos\theta} dEg(E)(1-f(E,E_F))=1, & \downarrow \end{cases} \quad (3)$$

For a transient state, the movement of the charge carriers could be described by Eq. (3). The demarcation energy, associated with the energy of the initial localized state, is determined by hopping downward of charge carriers at the moment (at which $v_l t = 1$). By connecting Eq. (1) and Eq. (3b), one has

$$\dfrac{[\ln(v_0 t)]^3}{24\alpha^3}\int_0^\pi d\theta \sin\theta \int_{-\infty}^{E_d(t)-\ln(v_0 t)\beta\cos\theta} dEg(E)(1-f(E,E_F))=1, \quad (4)$$

Moreover, considering that all of charge carriers move down towards the deep energy before the relaxation, here it is assumed that the probability of the empty site of the deep energy is 100%, that's, $1-f(E,E_F)=1$.

The charge carrier concentration can be determined as

$$n = \int_{-\infty}^{+\infty} dEg(E)f(E,E_d(t)), \quad (5)$$

Since the hopping of the charge carriers in the hopping space could result in the relaxation phenomena of the current and energy, the average hopping range $\overline{R(E)}$ is obtained by setting $N(T,\beta,E)=1$ in Eq. (2). Then, the diffusion constant $D(E)$ is calculated by the expression [16],

$$D(E) = \dfrac{\overline{R(E)}^2}{6(2\alpha)^2} v_0 \exp(-\overline{R(E)}), \quad (6)$$

The conductivity is calculated as [23]

$$\sigma_c = \int_{-\infty}^{+\infty} dE e^2 D(E) f(E,E_d(t)) g(E), \quad (7)$$

Finally, the current density in disordered systems is obtained as

$$J_n = \sigma_c \times F, \quad (8)$$

## 3. Results and discussion

Figure 1 shows the time-dependent of the demarcation energy at σ*=3. The parameters T=300 K, $N_t=1\times10^{28}$ m$^{-3}$, $\alpha^{-1}$=2.7 nm, F=3×10$^6$ V/m, $\nu_0$=1×10$^{12}$ s$^{-1}$, were used for the calculations. One can see that $E_d(t)$ decreases with the time increase, which suggests that the demarcation energy is far away from the shallower energy and shifts downward the deeper energy with time. Thus, it becomes more and more difficult for charge carriers to move away from a state by hopping with the time increase in disordered systems, due to the Gaussian distribution of DOS and the exponential-dependence rate on energy difference. The inset in figure 1 shows the temperature-dependent of charge carrier concentration at different time. The result indicates that the charge carrier concentration increases with the increase of the temperature, which reveals that the thermalizing effect is contributed to the increase of the free charge carrier concentration in disordered systems. The reason is that the charge carriers in deep states may move easily by thermal excitation to shallower states at higher temperature.

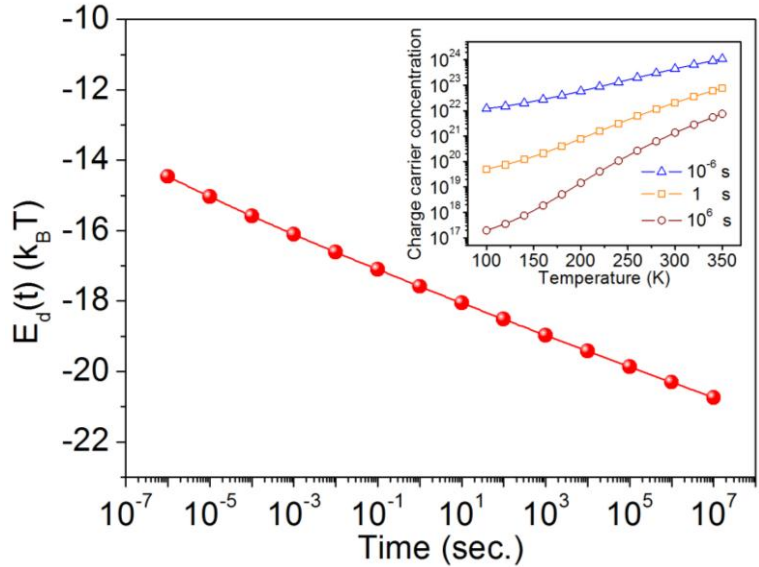

Fig. 1. The time-dependent of the demarcation energy at the disordered parameter σ*=3. Inset: the temperature-dependent of charge carrier concentration at different time.

Then, using the same parameters, we calculated the current density with time. Figure 2 shows the relaxation of current density with different disordered parameter σ as a function of the time. As can be seen that the current density of the disordered systems with the larger disordered parameter decreases faster than that with lower disordered parameter at a very short time, which indicates that the relaxation phenomena is more intense when the disordered parameter is lower in disordered systems. It is well known that charge carriers may travel through the material by hopping from one localized state to the other. The hopping rate at which this occurs is related to the conductivity of the disordered materials and the mobility. However, the material disorder can increase the charge carrier hopping barrier and hence decreases the conductivity and mobility of the disordered material, and eventually weakens the relaxation effect.

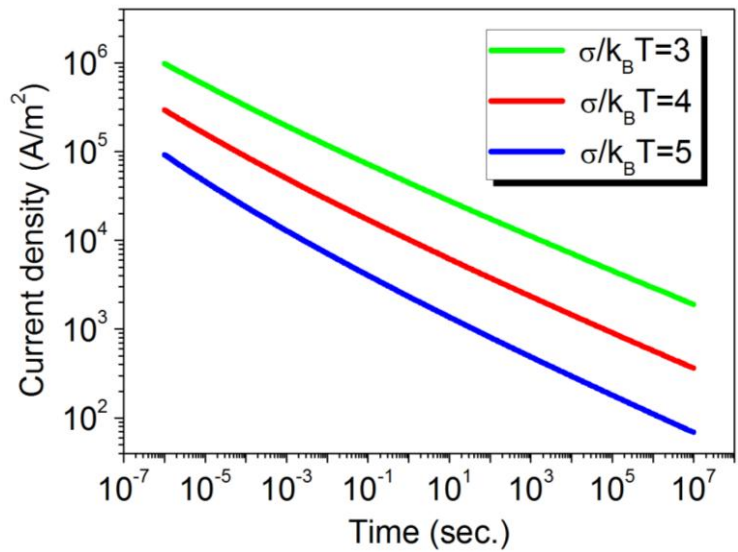

Fig. 2. The relaxation of current density with different disordered parameter σ as a function of the time.

Using the same parameters and keeping a constant disordered parameter ($\sigma/k_BT=4.5$), the electric field dependence of the current density with time was calculated, as shown in figure 3. It can be seen that the decrease of current density is slower in the higher electric field regime than that in the lower electric field regime, which reveals that the electric field may improve the relaxation effect. The reason for the increase of the relaxation rate is that, the electric field can change the energy difference among the localized states, and hence assists charge carrier hops along the electric field direction, which strengthens the relaxation rate of the current density.

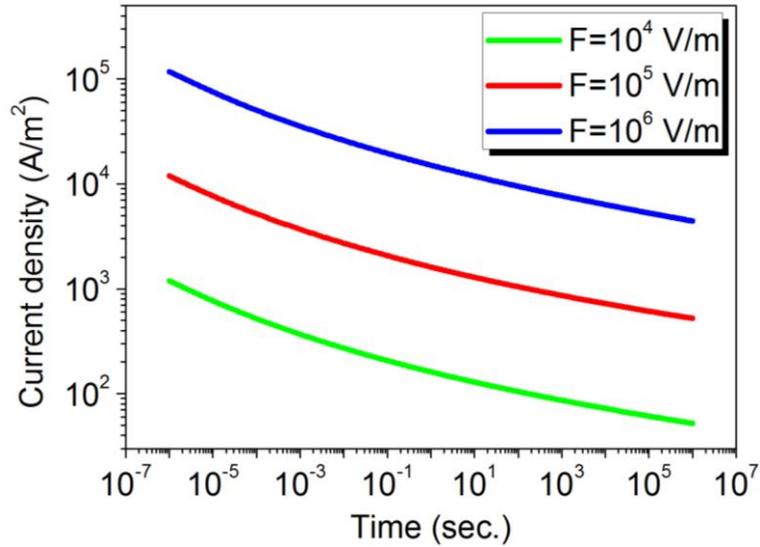

Fig. 3. The relaxation of current density with different electric field as a function of the time.

Lastly, we calculated the temperature dependence of the current density with time, shown in figure 4. The input parameter are σ= 0.078 eV, $N_t$=1×10$^{28}$ m$^{-3}$, $α^{-1}$=2.7 nm, F=3×10$^6$ V/m, $ν_0$=1×10$^{12}$ s$^{-1}$. It can be seen that the decrease of the current density at the lower temperature is faster than that at the higher temperature, which indicates that the relaxation effect can be strengthened at the higher temperature. As demonstrated above, the higher the temperature, the larger the free charge carrier concentration. Because the relaxation of the current and energy is attributed to the free charge carrier transport through localized states, therefore, the higher temperature in the systems can promote the relaxation effect in disordered systems.

As discussed above, the relaxation effect in disordered systems is mainly influenced by the disordered parameter, electric field and temperature, and hence our model strictly speaking is adequate for the different disordered systems. It is worth attention that, since the relaxation rate, corresponding to the slope value of the current density vs. time curve in the present work, is apparently attributed to the characteristic time τ when the stretching index β is a given value in the Kohlrausch law [6], our model can explain well the stretched-exponential law because the relaxation rate can be controlled by adopting the appropriate parameters, such as disordered parameter, electric field and temperature.

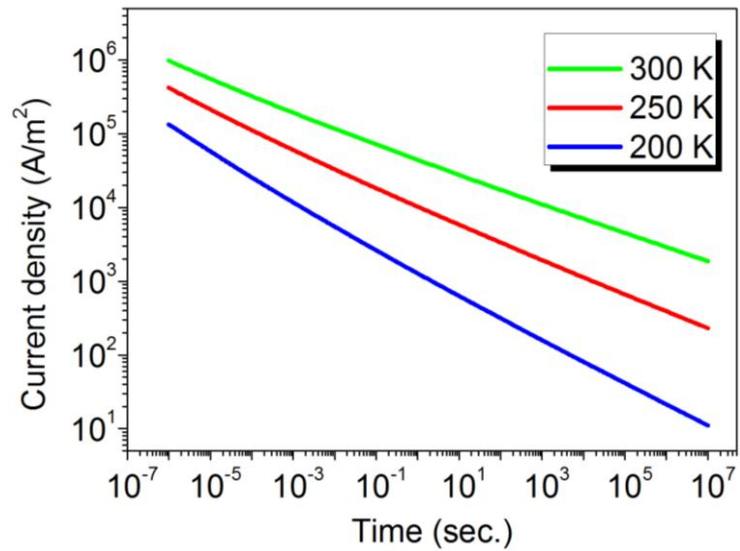

Fig. 4. The relaxation of current density with different temperature as a function of the time.

With the model discussed above, the relaxation of the current with time was calculated. The comparison between our work and experimental data [24] is shown in figure 5. The experimental result was measured in the Au/PPV/Au (PPV: phenylene-vinylene) device at room temperature with a Keithley 616 digital electrometer. One can see that our calculation result is agreed well with the experimental data except the initial value. The reason for the difference of initial current between the experiment and our model is that we ignore the downward movement process of charge carriers before the relaxation. The fitting parameter are $\sigma= 0.13$ eV, $N_t=1\times10^{28}$ m$^{-3}$, $\alpha^{-1}=1.25$ nm, $F=1.75\times10^6$ V/m, $\nu_0=1\times10^{12}$ s$^{-1}$, A (the device area) = 11 μm$^2$.

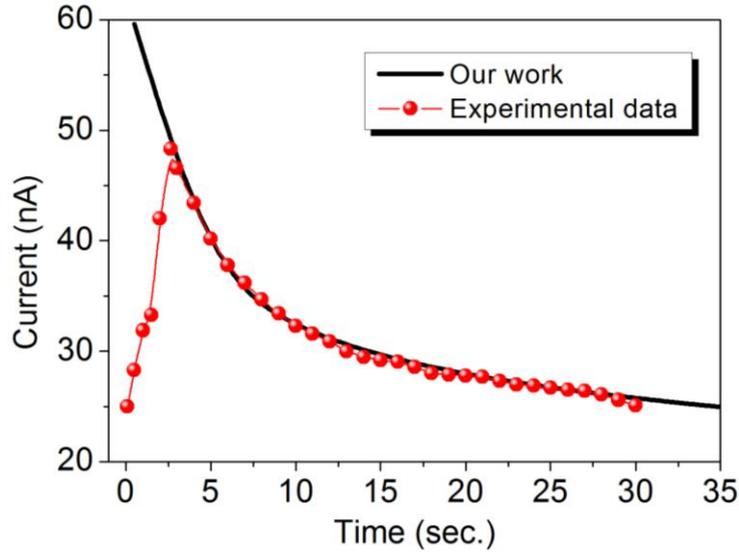

Fig. 5. Comparison between our work and experimental data.

## 4. Conclusion

We have proposed a model to explain the relaxation phenomena in disordered systems based on the hopping transport mechanism. The relaxation phenomena were discussed by applying the material disorder, electric field and temperature in disordered systems. The calculated results show that the material disorder can weaken the relaxation effect due to increase the charge carrier hopping barrier of the disordered systems. While the increase of electric field can promote the relaxation effect is the reason that the electric field can change the difference of the energy among the localized states and hence assists charge carrier jumps along the field direction. The temperature can strengthen the relaxation effect is attributed to the increase of the free charge carrier concentration. Our model can explain well the stretched-exponential law by adopting the appropriate parameters. The comparison to the experimental data for the organic semiconductors shows the proposed model is reliable.

**Acknowledge**

This research is supported by NSFC (No. 60825403, 61221004, 61274091, 61106119, 61106082, 60976003 and 61006011) and National 973 Program 2011CB808404.